\documentclass[]{emulateapj}
\usepackage{graphicx, amsmath, amsthm, amssymb,color,ulem}
\usepackage{graphics}
\usepackage{natbib}
\usepackage{ulem}
\def\in{_{\rm in}}
\def\out{_{\rm out}}
\def\b{_{\rm b}}
\def\c{_{\rm c}}
\newcommand{\evec}{\mathbf{e}}
\newcommand{\jvec}{{\mathbf{j}}}

\usepackage{amsmath, amsthm, amssymb,color}
\newcommand{\ba}{\begin{eqnarray}}
\newcommand{\ea}{\end{eqnarray}}

\def\red#1 {\textcolor{red}{#1}\ }   
\def\blue#1 {\textcolor{blue}{#1}\ }   

\shorttitle{Secular Transport  during disk dispersal}
\shortauthors{Petrovich, Wu \& Ali-Dib}

\begin{document}

\title{Secular transport during disk dispersal: the case of Kepler-419}
\author{Cristobal Petrovich\altaffilmark{1,2}, Yanqin Wu\altaffilmark{3} \& Mohamad Ali-Dib\altaffilmark{2}} 
\altaffiltext{1}{Canadian Institute for Theoretical Astrophysics, University of Toronto, 
60 St George Street, ON M5S 3H8, Canada; cpetrovi@cita.utoronto.ca}
\altaffiltext{2}{Centre for Planetary Sciences, Department of Physical \& 
Environmental Sciences, University of Toronto at Scarborough, Toronto, 
Ontario M1C 1A4, Canada}
\altaffiltext{3}{Department of Astronomy and Astrophysics, University of Toronto, ON M5S 3H4, Canada}

\begin{abstract}
Due to fortuitous circumstances, the two giant planets around Kepler-419 have well characterized 3-D orbits.  They are nearly coplanar to each other; the inner one has a large eccentricity ($\simeq 0.82$); and the apses of the two orbits librate around anti-alignment. Such a state defies available proposals for large eccentricities. We argue that it is instead uniquely produced by a decaying protoplanetary disk. When the disk was massive, its precessional effect on the planets forced the two apses to center around an anti-aligned state. And as the disk is gradually eroded, the pair of planets are adiabatically transported to a new state where most of the eccentricity (or rather, the angular momentum deficit) is transferred to the inner planet, and the two apses are largely anti-aligned. During this transport, any initial mutual inclination may be reduced or enhanced; either may be compatible with the current constraints.  
So a primordial disk can drive up planet eccentricities both in resonant planet pairs (as has been shown for GJ 876) and in secularly interacting, non-resonant pairs. The mechanism discussed here may be relevant for forming hot Jupiters and for explaining the observed eccentricities of warm and cold giant planets.  
\end{abstract}

\section{Introduction}

Orbital architectures of extra-solar planets give valuable information about their formation. Here, we focus on the system of Kepler-419. This system is remarkable in a number of ways. 

  First, one of the two giant planets (planet b) was discovered during transit, while its companion (planets c) was discovered through transit-timing variations \citep{dawson2012,ford12}. Both planets have been confirmed by radial velocity measurements and by modeling this data along with the Kepler photometry have yielded  three-dimensional  configuration of the system
  \citep{dawson2014}. Recent follow-up radial observations by  \citep{alme2018} have refined the orbital elements of the planets (summarized in Table \ref{table:TF}). Only perhaps one or two other systems have been so thoroughly characterized (Kepler-108 is a recent example, \citealt{MF2017}).  Second, the observed configuration is remarkable.  The two giant planets are nearly coplanar (mutual inclination $i_{\rm m}\lesssim10^\circ$).  They have a large semi-major axis ratio ($a\c/a\b\simeq4.5$, or period ratio $P\c/P\b\simeq9.7$), so their gravitational perturbations on each other are largely secular in nature.  The innermost planet has a high eccentricity ($e\b\simeq0.82$), while the outer one is mildly eccentric. Currently the apses appear nearly anti-aligned ($\varpi_b-\varpi_c\sim180^\circ$) and secular integrations indicate that the apses librate around anti-alignment with a small amplitude \citep{dawson2014,alme2018}. The relative longitudes of the ascending nodes is reported to librate around alignment ($\Omega_b-\Omega_c\sim0$) \citet{alme2018}, but this is merely the result of projection and not a true dynamical effect. The apse anti-alignment means the system lies near one of the two secular fixed points (the other being aligned), with most of the angular momentum deficit (AMD) in the system retained in the inner planet. Such a configuration begs the question of its origin.

The fortuitous viewing geometry of Kepler-419 affords us a rare
 glimpse of its 3-D geometry, but such a system may not be uncommon.

\subsection{Failed explanations for Kepler-419 and this work}

Here, we discuss and reject a number of explanations for such a system, before introducing our proposed scenario.

The high eccentricity of planet b is evocative of the so-called Lidov-Kozai mechanism \citep{kozai,lidov}. However, this requires the presence of a companion on a significantly inclined orbit, opposite to what is observed for planet c. 

Planet-planet scattering from an initially more populous system is a possibility. However, there are several properties of the system that disfavor this scenario. In descending order of strength at constraining the scattering scenario:
\begin{enumerate}
\item The libration around a high-eccentricity and anti-aligned fixed point is an unlikely outcome of scattering. Although scattering can produce a significant fraction of systems with librating apsidal configurations, the large majority does so around alignment \citep{BG06,BG07}. In order to quantify the distance of two-planet systems to the libration/circulation boundary after scattering,  \citet{timpe2013} introduced the following quantity
\ba
\epsilon=\frac{2\min \sqrt{x^2+y^2}}{\left(x_{\rm max}-x_{\rm min}\right)+\left(y_{\rm max}-y_{\rm min}\right)},
\ea
where $x=e\b e\c \sin(\varpi\b-\varpi\c)$ and $y=e\b e\c \cos(\varpi\b-\varpi\c)$. The maximum and minimum are computed after several secular cycles, and \citet{timpe2013} finds $\epsilon<2$ for $>95\%$ of their scattering experiments (figure 2 therein). In stark contrast, Kepler-419 is far from the libration/circulation boundary boundary and has $\epsilon\sim3$.

\item The near coplanarity of the planets is an unlikely outcome from scattering exciting such large eccentricities \citep{JT2008,chatt2008}.  In particular, \citet{Simbulan2017} finds that mutual inclinations follow a Rayleigh distribution with median $\simeq30^\circ$.
\item There is a tendency for planet-planet scattering to lead to equipartition of AMD as well as mass segregation (more massive planet inside, e.g., \citealt{chatt2008,JT2008}), while Kepler-419 is far from AMD equipartition (the inner planet's AMD is $\simeq 4.4$ larger than that of the outer planet) and the outer planet is nearly three times more massive.
\end{enumerate}
The proximity of the planet pair to a secular fixed point suggests a dissipative or an adiabatic process in the past (e.g., \citealt{malhotra2002}). If the inner planet experiences damping (e.g., through tidal interaction with the star), past studies \citep{WG2002,zhang2013} have shown that the anti-aligned mode (in the context of the classical Lagrange-Laplace secular theory), in which the inner planet is typically more eccentric, is damped more rapidly than the aligned mode when $m\b a\b^{1/2}<m\c a\c^{1/2}$, driving the system towards the fixed point of apse alignment.
\citet{alme2018} designed an initial state with the pair starting near its current state ($e_b \simeq 0.9$) but with a larger libration amplitude around the anti-alignment fixed point. They show that dissipation on the inner planet may reduce this libration amplitude to the observed value, while avoiding to damp the inner eccentricity so dramatically that the planet is dragged in to become a hot Jupiter. However, such an initial state is so close to the current one, it begs the question of how itself could have arisen naturally.
 
This leaves adiabatic transport as the remaining candidate. This refers to the case where the system is adiabatically transported to the current state from a generic initial state, due to a gradual change in the environment. Here, we propose that a particular environmental change, a slowly decaying proto-planetary disk, can neatly and naturally lead to the current configuration.

Such a proposal does not require an overly imaginative set-up. exoplanets are formed in disks and these disks have by now long vanished. But the dynamical impacts of the disk are likely permanently imprinted in the planetary orbits \citep[e.g.][]{goldreich,KN2012}. Well-known examples include the GJ 876 planet pair that is possibly pushed into mean-motion resonances by the now extinct disk \citep{LP86,LP2002}. Here, we focus on one specific dynamical effect, secular precession of the planet orbit driven by a massive disk. We are able to show that, as the precession wanes away, a planet pair will be securely transported to an anti-aligned state with most of the AMD transferred to the inner planet.

 Later on, we further argue that this may have implications in a much broader context.

\begin{table}
  \caption{Parameters of Kepler-419 planets}
  \centering
\begin{tabular}{lll}
\hline  
\hline
 &planet b &  planet c  \\ 
  \hline 
mass $m$ $[M_J] $& $2.77\pm 0.19$ & $7.65\pm0.27$\\
semi-major axis $a$ [AU]& $0.3745\pm 0.0046$ & $1.697\pm 0.02$\\
eccentricity $e$ & $0.817\pm 0.016$ & $0.1793\pm 0.0017$\\
inclination $i$ [deg] & $87.04\pm 0.72$ & $87\pm2$\\
arg. of pericenter $\omega$ [deg] & $94\pm 2.2$ & $275.7\pm1.8$\\
long. asc. node $\Omega$ [deg] & $180$ (fixed) & $185.4\pm7.6$\\
  \hline \hline
\end{tabular}
\\ {These values correspond to the median and $68.3\%$ confidence interval (Table 2 of \citet{alme2018}). The mass of the host star is $M_s=1.438\pm 0.053M_\odot$. The $68.3\%$ confidence interval of the mutual inclination between planets b and c is $i_{\rm m}=[1.2^\circ, 7.58^\circ]$, based on secularly evolved posteriors distributions.
  \\ 
  }\label{table:TF}
\end{table}

\subsection{`Secular Resonance Sweeping'}

Similar dynamics as the one we are invoking has been investigated in different contexts. In particular, the term `secular resonance sweeping' has often been used to describe this process, bringing with it a certain amount of confusion. Secular resonance refers to the case where the secular frequencies (or their linear combinations) in a system are commensurate with each other \citep[see, e.g.][]{laskar,LW2011,BMH2015}. In our solar system, some orbits are precessed by the planets at rates that are comparable to the planets' own precession rates. Here, a test particle's eccentricity can be strongly excited and the associated secular angle librates around $0$ or $180^\circ$. In the early Solar system, a decaying disk could have altered the secular frequencies in a time-dependent way, causing locations of secular resonances to sweep across the system, possibly exciting eccentricities and inclinations of asteroids and Mars \citep{ward81,Hepp1980}. Equivalently, the wanning quadrupole moment of the Sun as it spins down can affect objects in the very inner region \citep{ward76}.

Such an idea has found applications in the exoplanetary field, e.g., eccentricity variations 
and apsidal locking  of radial velocity exoplanets \citep{naga2003,moeckel2008}, eccentricity excitation of debris disks (or small planets) perturbed by outer planets in a dispersing nebula \citep{BK2017,zheng2017} or a migrating planet \citep{MM2009,MM2011}; eccentricity and inclination excitation of two-planet systems induced by a decaying host star's quadrupole moment \citep{BBL2016,SB2017,SMB2018}; inclination excitation of planets in binaries with disks \citep{LM2016,martin2016,MK2017,ZL2018}. 

The dynamics we invoke here is subtly different from most of these examples. The easiest way to visualize our mechanism is to think in the frame-work of the so-called `Laplace-Lagrange' theory \citep[e.g.][]{MurrayDermott} where the secular perturbations are analyzed to the linear order in the eccentricity. Interactions between a pair of planets give rise to two fixed points, one with aligned apses, the other anti-aligned. When the precession induced by the disk greatly overcomes that of the planet-planet interactions, only one mode survives: generally\footnote{This statement holds (at least) when the circular angular momentum of the inner planet is smaller than of the outer (like in Kepler-419) and the disk leads to an effective retrograde precession ($\dot{\varpi}\b-\dot{\varpi}\c<0$).} the anti-aligned mode.
Systems close to this point will remain close to this fixed point, as environments are slowly modified.
This is more aptly called `adiabatic transport', as opposed to `resonance capture', as there is no need for the presence of a separatrix.
Essential features of this dynamics are retained in the nonlinear order \citep{BBL2016}, and even when inclinations are considered, as we show here.

Many of the features of the adiabatic transport driven by a decaying disk have been previously discussed in the context of exoplanets by \citet{naga2003}. However, these authors focused on other systems (e.g., $\upsilon$ Andromedae), for which the imprint of this mechanism remains ambiguous \citep{Deitrick2015}.

\section{Model}

We describe a simple model to understand how  a two-planet system like Kepler-419 evolves secularly during the gas disk dispersal phase. 

We make two key assumptions about the initial states of the system:
\begin{itemize}
\item there was a massive disk outside the planetary orbits. It gradually decays away with time. We only consider the precessional effect of such a disk on the planets, and ignore issues like migration, eccentricity and inclination damping.
\item there was a moderate amount of AMD and it was mostly concentrated in the outer planet. 
\end{itemize}
We discuss the needs for and the validities of these assumptions in \S \ref{sec:discussion}.

In our notation, the inner planet b has a mass $m\b$, orbital period $P\b$, semi-major axis $a\b$, eccentricity $e\b$, argument of pericenter $\omega\b$, longitude of the ascending node $\Omega\b$, and longitude of pericenter $\varpi\b=\omega\b+\Omega\b$. The same for the outer planet c.  The currently measured values for Kepler-419 from \citet{alme2018} are shown in Table 1.

Besides from the above orbital elements, we shall also use the vectorial orbital elements, where the eccentricity vectors are ${\bf e}\b$ and ${\bf e}\c$, and the dimensionless orbital angular momentum vectors are defined as ${\bf j}\b=(1-e\b^2)^{1/2}{\bf \hat{j}\b}$ and  ${\bf j}\c=(1-e\c^2)^{1/2}{\bf \hat{j}\c}$, with hats indicating unit vectors  (see, e.g, \citealt{TTN09}).

In the following, we introduce the relevant secular interactions, between planet-planet, and planet-disk.
For the vectorial elements the interaction energy will be denoted as potential $\phi$ and for the standard orbital elements as a Hamiltonian $H$.  

\subsection{Secular planet-planet interactions}

Given the large ratio of semi-major axis ($a_c/a_b \simeq 4.5$), we expand the interaction energy using $\alpha = a_b/a_c$ as a small parameter. This is exact to all eccentricities and inclinations.  The doubly time-averaged interaction potential up to octupole order ($\alpha^3$) can be written in dimensionless form as (e.g., \citealt{liu15,petro15}):  
\ba &\phi_{\rm{planet}}&= -\frac{\phi_0}{(1-e\c^2)^{3/2}} \Big[ \tfrac{1}{2} \big({\bf j}\b\cdot {\bf \hat{j}}\c\big)^2 + \left(e\b^2 -\tfrac{1}{6}\right)-
\tfrac{5}{2}  \big({\bf e}\b\cdot {\bf \hat{j}}\c\big)^2   \Big]\nonumber\\
&-& \frac{25a\b\phi_0}{16a\c\left(1-e\c^2\right)^{5/2}} \Big\{
 \big({\bf e}\b\cdot {\bf e}\c\big)
 \Big[ \big(\tfrac{1}{5}-\tfrac{8}{5}e\b^2\big)
 - \big({\bf j}\b\cdot {\bf \hat{j}}\c\big)^2
\nonumber\\
 &+& 
 7  \big({\bf e}\b\cdot {\bf \hat{j}}\c\big)^2
\Big]
- 
2\big({\bf j}\b\cdot {\bf \hat{j}}\c\big)
\big({\bf e}\b\cdot {\bf \hat{j}}\c\big)
 \big({\bf j}\b\cdot {\bf \hat{e}}\c\big)
\Big\},
	 \label{eq:phi_oct}
\ea
where
   \ba
   \phi_0&=&\frac{3Gm\b m\c a\b^2}{4a\c^3}\, ,
\label{eq:phi_0}
\ea
represents the magnitude of quadrupole interactions. Scaling this term by the monopole stellar potential, we obtain the characteristic secular timescale for the inner planet as \citep[also referred to as the Lidov-Kozai timescale, e.g.,][]{naoz2016}:
\ba
\tau_{\rm sec,b}\equiv \frac{m\b\sqrt{GM_sa\b}}{\phi_0}
=\frac{2P\b}{3\pi}\frac{M_{\rm s}}{m\c}\frac{a\c^3}{a\b^3}\simeq 710\mbox{ yr}
\label{eq:tau_sec},
\ea
while for the outer planet it becomes
\ba
\tau_{\rm sec,c}\equiv  \frac{m\c\sqrt{GM_sa\c}}{\phi_0}
=\tau_{\rm sec,b}\frac{m\c a\c^{1/2}}{m\b a\b^{1/2}}\simeq 4.2\times10^3\mbox{ yr}
\label{eq:tau_sec_c}.
\ea
We explored the effect of higher-order terms (hexadecupole, up to $\alpha^4$) and found these to be unimportant, as is expected.

\subsection{Secular perturbations from a massive disk}

We consider precessional effect by a flat disk. We introduce
 an axisymmetric disk with a truncated density profile
\ba
\Sigma(R)=\Sigma_0 \left(\frac{R}{R\in}\right)^{-\gamma}~~ \mbox{for}~~R\in<R<R\out,
\label{eq:sigma_r}
\ea
and $\Sigma(R)=0$ elsewhere. The disk has a mass
\ba
M_{\rm disk}=2\pi\Sigma_0 R\in^2\left[\frac{\left(R\out/R\in\right)^{2-\gamma}-1}{2-\gamma}\right].
\ea

In the Appendix we have derived the potential from this density profile averaged over the orbit of an interior planet $p=\{{\rm b,c}\}$, $a_p<R\in$, for any value of $\gamma$ and $R\out/R\in$ (Equation \ref{eq:phi_d}). For $\gamma=1.5$ and $R\out/R\in\gg1$, this potential reads
\ba
\phi_{\rm disk}&=&-\frac{3Gm_p M_{\rm disk} a_p^2}{20R\in^3}\left(\frac{R\in}{R\out}\right)^{1/2} \mathcal{B}
\times \left[e_p^2+ {\bf j}_p\cdot \hat{\bf n}_{\rm d}\right],
\label{eq:phi_d_1.5}
\ea
where $\mathcal{B} = \mathcal{B}\left(\gamma=1.5,R\out/R\in,a_p/R\in\right)$  is a factor of order unity to correct for the contribution from regions of disk that are close to the planet (Eq. \ref{eq:B}). In Figure \ref{fig:beta} we observe that $\mathcal{B}$ depends weakly on the density profile slope $\gamma$ and the radial extent $R\out/R\in$, while it increases as the planet approaches the inner edge of the cavity, $a_p/R\in\to 1$. If $R_{\rm in}=1.5a_p$, we have $\mathcal{B}\simeq2$, while for $R_{\rm in}=2a_p$ we have $\mathcal{B}\simeq1.4$.

The corresponding precession rate in units of mean motion, $n_p=(GM_s/a_p^3)^{1/2}$,  is simply given by the ratio of the disk potential and the stellar monopole resulting in the following precession timescale:
\ba
\tau_{{\rm disk},p}=\frac{10}{3\pi \mathcal{B}}
\frac{M_s}{M_{\rm disk}}
\frac{R\in^3}{a_p^3}
\left(\frac{R\out}{R\in}\right)^{1/2}
P_p.
\label{eq:tau_disk}
\ea
As a reference, for planet c with $R\in\sim1.5a\c$ and $R\out/R\in=10$, the precession timescale becomes
\ba
\tau_{{\rm disk}, c}\simeq105\mbox{ yr}
\left(\frac{0.1M_s}{M_{\rm disk}}\right).
\ea

 From Hamilton's equations the angles relevant for subsequent analysis vary as
\ba
\frac{d\varpi_p}{dt}&=&\frac{2(1-e_p^2)^{1/2}-1}{\tau_{{\rm disk},p}}=\frac{1}{\tau_{{\rm disk},p}}\left[1+\mathcal{O}(e_p^2)\right],
\label{eq:vpi_dot}\\
\frac{d\Omega_p}{dt}&=&-\frac{1}{\tau_{{\rm disk},p}}.  \ea

In the following, we apply these elements to study the secular evolution under three cases: first two simplified cases, coplanar but eccentric, circular but non-coplanar, and lastly, the realistic situation of eccentric and non-coplanar orbits. The simplified cases are useful to understand the full case. 

\begin{figure}
\center
\includegraphics[width=8.8cm]{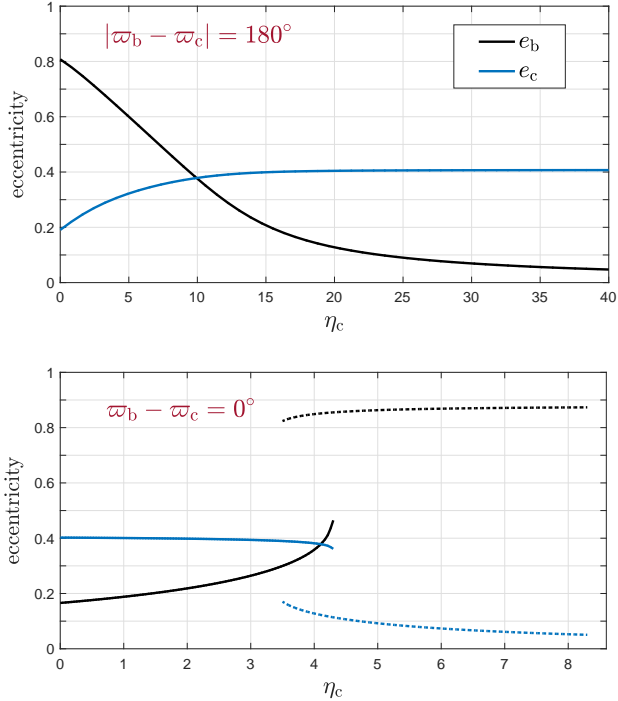}
\caption{
Fixed points of the Hamiltonian in Equation (\ref{eq:H_cop}) for different values of the parameter $\eta\c\propto M_{\rm disk}$ (Eq. [\ref{eq:eta_c}]). The total angular momentum is set to be the current day value (Eq. [\ref{eq:J_tot}]).
Top and bottom panels show the branches with $|\varpi\b-\varpi\c|=180^\circ$ and  $\varpi\b-\varpi\c=0^\circ$, respectively.
In our model, as $\eta_c$ is adiabatically decreased from the right to the left, following the upper fixed point will lead to eccentricity (or rather AMD) being transferred from the outer to the inner planet. The reverse occurs if one follows the fixed points in the lower panel.
Note the two panels have different horizontal scales. No aligned fixed point exists for $\eta_c \geq 8.5$.}
\label{fig:fixed_point}
\end{figure}

\begin{figure*}
\center
\includegraphics[width=18cm]{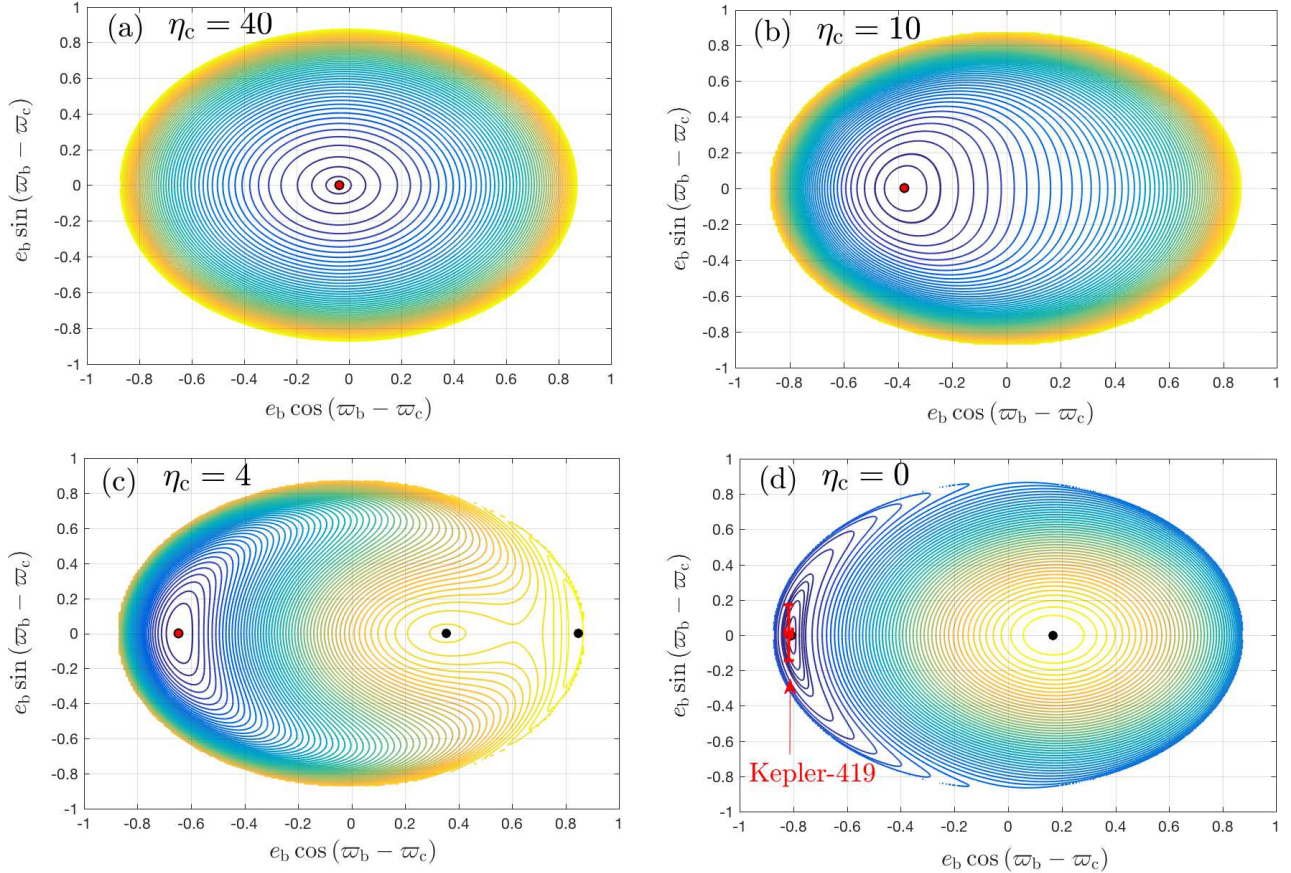}
\caption{Contours of constant Hamiltonian (Equation \ref{eq:H_cop}) in the plane of the (pseudo) coordinate-momentum pair $e\b[\cos (\varpi\b-\varpi\c),\sin (\varpi\b-\varpi\c)]$ 
 and for different importance of the disk, represented by the different values of $\eta\c \propto M_{\rm disk}$ (Eq. [\ref{eq:eta_c}]).
The orbital angular momentum is  chosen to match the current value for Kepler-419 (Eq. [\ref{eq:J_tot}]).
 Fixed points from Equation (\ref{eq:fixed_point}) are shown as filled red circles.  
The color indicates energy with bluer corresponding to lower values. 
At large disk mass ($\eta_c \gg 1$), only the anti-aligned fixed point exists and all trajectories librate or circulate around this point. At $\eta_c$ decreases,  this  fixed  point  moves  further  to  the left and , two new, aligned, fixed points appear: one with moderate $\b$ and the other with large $e\b$. By the time $\eta_c = 0$, the aligned fixed point with moderate $e\b$ commands most of the phase space, with the original anti-aligned fixed point pushed to the far left (high value of $e_b$). 
In the bottom-right panel ($\eta_c = 0$), the current state of Kepler-419 is marked by a red point, together with its $1-\sigma$ error bars.
}
\label{fig:contour_ecc}
\end{figure*}

\section{Coplanar case: anti-alignment and eccentricity transfer}
\label{sec:apsidal}

In this section we assume that all three orbital planes (for the two planets and the disk) coincide. 

In this case, ${\bf e}\b\cdot {\bf e}\c=e\b e\c\cos (\varpi\b-\varpi\c)$ and ${\bf e}\b\cdot {\bf j}\c=0$, and the full Hamiltonian (Eqs. [\ref{eq:phi_oct}] and [\ref{eq:phi_d_1.5}]), can be reduced to
\ba
&&\tilde{H}_{\rm cop}=-\eta\b\big[e\b^2+(1-e\b^2)^{1/2}\big]-
\eta\c\big[e\c^2+(1-e\c^2)^{1/2}\big]\nonumber\\
&&-\frac{(e\b^2/2+1/3)}{(1-e\c^2)^{3/2}}
+\frac{5a\b}{16a\c}\frac{4+3e\b^2}{(1-e\c^2)^{5/2}}e\b e\c\cos\left(\varpi\b-\varpi\c\right),\nonumber\\
\label{eq:H_cop}
\ea
where we have divided both sides by the constant $\phi_0$ (Eq. \ref{eq:phi_0}). The dimensionless factors $\eta$ measure the relative importance of the secular effect from the disk, compared to that from the other planet.
For $\gamma=1.5$ and $R\out/R\in\gg1$, we have
\ba
\eta\b&=&\frac{\mathcal{B}}{5}\frac{M_{\rm disk}}{m\c}\left(\frac{a\c}{R\in}\right)^3\left(\frac{R\in}{R\out}\right)^{1/2},\\
\eta\c&=&\frac{\mathcal{B}}{5}\frac{M_{\rm disk}}{m\b}\left(\frac{a\c}{R\in}\right)^3
\left(\frac{a\c}{a\b}\right)^2\left(\frac{R\in}{R\out}\right)^{1/2}.
\label{eq:eta_c}
\ea
For Kepler-419, we have
\ba 
\frac{\eta\c}{\eta\b}=\frac{m\c a\c^2}{m\b a\b^2}\simeq 57\, .
\ea
So one could ignore the precession of the inner planet due to disk ($\eta_b$)  without losing any significant dynamical effect.
In the following, we only specify the value of $\eta\c$, although we still retain the small contribution from $\eta\b$.

For reference,  setting $R\out/R\in=10$, $\gamma=1.5$, $R\in /a\c=1.5$, we find that the disk is important (for planet c) when it is comparable in mass to the planets,
\ba
\eta\c \simeq 0.76 \times \frac{M_{\rm disk}}{m\b} \simeq 2.1\times\frac{M_{\rm disk}}{m\c}\, .
\label{eq:eta_approx}
\ea
The above coefficient decreases from $2.1$ to $\simeq 0.62$ if we moves the inner edge of the disk further to $R\in /a\c=2$.

As the external disk is axisymmetric, it does not exchange angular momentum with the planets and the pair's total orbital angular momentum is conserved. This can also be observed from the form of the Hamiltonian, where only the angle combination $\varpi_b - \varpi_c$ appears\footnote{One can adopt the following set of canonical variables for our analysis \citep{MM2004}:
  $\left[m_b \sqrt{a_b} (1- \sqrt{1-e_b^2}), \varpi_c - \varpi_b \right]$ and
$\left[m_b \sqrt{a_b} (1- \sqrt{1-e_b^2}) + m_c \sqrt{a_c} (1-\sqrt{1-e_c^2}),-\varpi_c \right]$
with the canonical momentum in the form of AMD (deficit relative to circular orbit). In this form, it is clear that the Hamiltonian contains only one degree of freedom (the first pair) and is integrable.}.
Thus,  we further specify that the total angular momentum is the same as current day value, 
\ba
\mathcal{J}=(1-e\b^2)^{1/2} \beta+(1-e\c^2)^{1/2}\simeq1.082\, ,
\label{eq:J_tot}
\ea
where 
\ba
\beta=\frac{m\b a\b^{1/2}}{m\c a\c^{1/2}}\simeq\frac{1}{6}
\label{eq:beta}
\ea
is the ratio between the angular momenta for circular orbits of the planet  b and c.

In Figure \ref{fig:fixed_point}, we show fixed points for the above Hamiltonian, calculated by setting $d(\varpi\b-\varpi\c)/dt=0$, which leads to the following algebraic condition from Hamilton's equations:
\ba
\frac{d\tilde{H}_{\rm cop}}{de\b}
\frac{(1-e\b^2)^{1/2}}{e\b}-
\beta\frac{d\tilde{H}_{\rm cop}}{de\c}\frac{(1-e\c^2)^{1/2}}{e\c}=0\, .
\label{eq:fixed_point}
\ea
This is subject to the constraint of Eq. (\ref{eq:J_tot}).
The fixed points consist of two branches: $|\varpi\b-\varpi\c|=180^\circ$ and $|\varpi\b-\varpi\c|=0^\circ$. We observe that for the fixed point at $|\varpi\b-\varpi\c|=180^\circ$ (panel a), $e\b$ always grows at the expense of decreasing $e\c$. The opposite behavior occurs for the fixed points at $|\varpi\b-\varpi\c|=0^\circ$ (panel b), which can manifest in in two branches: (i) a high $e\b$-branch for $\eta\c\sim3.5-8.5$; (ii) a low $e\b$-branch for $\eta\c\lesssim4.3$. 

In Figure \ref{fig:contour_ecc}, we further demonstrate the dynamics by plotting contours of  constant Hamiltonian (Eq. \ref{eq:H_cop}) in the plane of the cartesian (pseudo) coordinate-momentum pair\footnote{The canonical pair is actually $\{2m_b \sqrt{a_b} [1- (1-e_b^2)^{1/2}]\}^{1/2}
[\cos (\varpi\b-\varpi\c),\sin (\varpi\b-\varpi\c)]$, which for moderate values of $e\b$ is proportional to
$e\b[\cos (\varpi\b-\varpi\c),\sin (\varpi\b-\varpi\c)]$.}, $e\b[\cos (\varpi\b-\varpi\c),\sin (\varpi\b-\varpi\c)]$ for a few different values of $\eta\c \propto M_{\rm disk} $ (Eq. [\ref{eq:eta_c}]). The orbital angular momentum is fixed to be that in Equation (\ref{eq:J_tot}). 

When $\eta\c=40$, which is roughly equivalent to $M_{\rm disk}\sim20m\c$ or $\sim 0.1M_s$, only the anti-aligned fixed point is present, and contours of constant Hamiltonian appear as roughly concentric circles around the origin (actually slightly shifted to anti-aligned apsidal orientations). All initial configurations circulate around this fixed point. As $\eta_c$ decreases, this fixed point moves further to the left, and for $\eta_c \lesssim 8.5$ ($\eta_c \lesssim 4.5$), a new fixed point with large (moderate) $e_b$ and aligned apses appears on the right (see two fixed points panel c with $\eta\c=4$). When $\eta_c$ decreases to zero (disk vanishes), we see that most of the space is now taken up by trajectories around the aligned fixed point (the other aligned fixed point with large $e\b$ disappears), with the anti-aligned fixed point commanding a small set of curves that librate around $|\varpi_b - \varpi_c| = 180^\circ$.

\subsection{Aligned vs. anti-aligned fixed points}

At $\eta_c = 0$, the two fixed points (one aligned, one anti-aligned) are easily understood, at the low-eccentricity limit, using the classical Laplace-Lagrange theory \citep[cf][]{MurrayDermott}. The Hamiltonian equation can be transcribed into one that describes the interaction of two linear harmonic oscillators, with the general solution being the linear combination of two eigen-modes, one having aligned apses, the other anti-aligned. Typically, the eccentricity of the inner planet is lower than the outer one in the aligned mode, and higher in the anti-aligned mode. Much of these features are  retained when eccentricities are no longer small, as can be seen in Figure \ref{fig:fixed_point}.

What happens at $\eta_c \gg 1$? Why is there only one anti-aligned fixed point? This can be understood in the following simple analysis. We work at the low eccentricity limit. As the inner planet has a much smaller inertia, we assume it to be a test particle ($m_b\ll m_c$). So $e_c$ remains a constant, while its apsidal angle, $\varpi_c$ precesses uniformly under the disk's influence as $\varpi_c = t/\tau_{{\rm disk},c}$. Under these simplifications (that retain the essential dynamics), the Hamiltonian in Eq. (\ref{eq:phi_oct}) can be reduced, for the inner test particle, into a time-dependent form,
\begin{equation}
H_{\rm tp}= {\phi_0}\left[-\frac{e\b^2}{2}+\frac{5\alpha e\c}{4}  \, e\b\cos\left(\varpi\b-{{t}\over{\tau_{\rm disk,c}}}\right)\right]\, ,\nonumber\\
\label{eq:ham-tp}
\end{equation}
where $\alpha= a_b/a_c$.
To remove the time-dependence, we carry out a canonical transformation from an initial canonical pair $\{-\varpi\b,J\b=m\b\sqrt{GM_s a\b}(1-\sqrt{1-e\b^2})\}$ to a new variables $\{\varpi\b',J\b'\}$ using the following generating function \citep[e.g.,][]{MM2011}:
\ba
\mathcal{F}(-\varpi\b,J\b',t)=(t/\tau_{\rm disk,c}-\varpi\b)J\b',
\ea
so $\varpi\b'= \partial {\mathcal{F}}/\partial J_b'=t/\tau_{\rm disk,c}-\varpi\b$, and $J\b=-\partial {\mathcal{F}}/\partial \varpi_b=J\b'$. And the new Hamiltonian reads
\ba
H_{\rm tp}'&=&H_{\rm tp}+\frac{d\mathcal{F}}{dt}\nonumber\\
&=&\phi_0\left[-\frac{e\b^2}{2}+{{5 \alpha e_c}\over{4}} e\b
\cos\left(\varpi\b'\right)\right]\nonumber\\
&&+ \frac{m\b\sqrt{GM_s a\b}}{\tau_{\rm disk,c}}\left[1-(1-e\b^2)^{1/2}\right],
\ea
which can be expressed in a dimensionless form as 
\ba
\tilde{H}_{\rm tp}'&\simeq& (\eta -1) \frac{e\b^2}{2}+ {{5 \alpha e_c}\over{4}} e\b
\cos\left(\varpi\b'\right)
\ea
where the dimensionless parameter $\eta$ scales the disk precession rate on planet c by the precession rate on b by c,
\ba
\eta=\frac{\tau_{\rm sec,b}}{\tau_{\rm disk,c}}= \beta \eta_c\, .
\ea

 When the disk is absent, ($\eta_c = \eta = 0$), the fixed point lies at $\varpi\b' = 0$ with $e_b$ secularly forced to $e_{\rm b,eq} = 5/4 \, \alpha e_c$. This is the usual solution of a test particle secularly forced by an external eccentric planet. However, when $\eta_c \gg 1/\beta \sim 6$, planet c precesses so rapidly by the disk, in the frame that rotates with its apse, the equilibrium position for planet b is anti-aligned ($\varpi_b' = \pm \pi$) with 
\begin{equation}
e_{\rm b,eq} \approx {5\over 4}\, {{\alpha e_c}\over{\eta-1}} \, .
\label{eq:eb_anti}
\end{equation}
This anti-aligned fixed point moves to higher eccentricity as $\eta = \eta_c \beta$ approaches unity, at which time nonlinear terms will need to be included to describe the evolution.

\subsection{Adiabatically changing the disk mass}

Next, we consider the evolution when the gas disk slowly disperses. In particular, we assume that the parameter $\eta_c =\eta\c(t)$ is started at a large value and decreases adiabatically, meaning that the gas disperses on timescales much longer than the libration/circulation timescale dictated by the secular planet-planet interactions $\tau_{\rm sec,b}$ (Eq.  [\ref{eq:tau_sec}]), or,
\ba
\left|\frac{d\log M_{\rm disk}}{dt}\right|^{-1}\gg \tau_{\rm sec,b}\sim10^3{\rm yr}.
\ea
This is a reasonable assumption as the dispersal timescales of protoplanetary disks are typically of order $\sim1$ Myr \citep{reviewpe}. 

The theory of adiabatic invariants (see, e.g., \citealt{goldstein}) indicates that the dynamics of the system largely follows that in Figure \ref{fig:contour_ecc} in the secular time-scale, but gradually shifts from one panel to the next panel as the value of $\eta_c(t)$ decreases, keeping the canonical phase space volume constant.
Consider a system initially started near the anti-aligned fixed point in panel a of Figure \ref{fig:contour_ecc}. This is equivalent to stating that $e_b$ is initially small. As $\eta_c$ decreases, the system will be transported, along with the anti-aligned fixed point, to the left (larger $e_b$). By the time $\eta_c = 0$, the planet pair will be librating around the anti-aligned fixed point at very high $e_b$, exactly what is observed for Kepler-419 today. 

In conclusion, if the inner planet starts with small  eccentricity (and outer one with a moderate value)  adiabatic removal of the disk potential will transfer eccentricity (or more precisely, AMD) from the outer to the inner planet, driving the relative apsidal angle to anti-alignment.
The end-state of this process naturally reproduces the orbital configuration of Kepler-419.
In \S \ref{sec:evol}, we conduct numerical integrations to confirm this picture. 

 \begin{figure}
\center
\includegraphics[width=8.7cm]{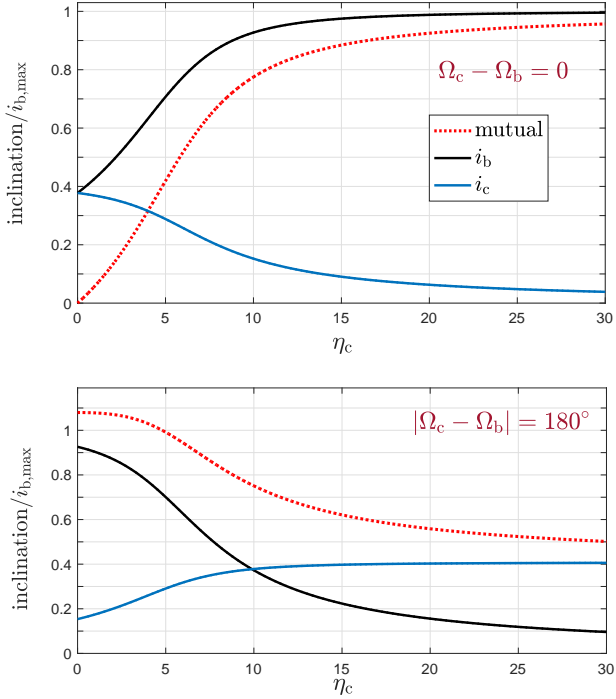}
\caption{Evolution of the inclinations following the fixed points of the Hamiltonian in Equation (\ref{eq:H_inc}) as a function of the adiabatically decreasing parameter $\eta\c\propto M_{\rm disk} (t)$ (Eq. [\ref{eq:eta_c}]). The upper panel shows the fixed point at  $\Omega\b-\Omega\c=0$  (Eq. [\ref{eq:Delta_Om_0}]), where the evolution leads to a decrease of the mutual inclination (red dashed line). The lower panel shows the fixed point at $|\Omega\b-\Omega\c|=180^\circ$  (Eq. [\ref{eq:Delta_Om_pi}]), where the evolution leads to an increase of the mutual inclination. The inclinations are given in units of the maximum inclination attainable by planet b from Equation (\ref{eq:ib_max})}
\label{fig:i_fixed}
\end{figure}
 
  \begin{figure*}
\center
\includegraphics[width=18.5cm]{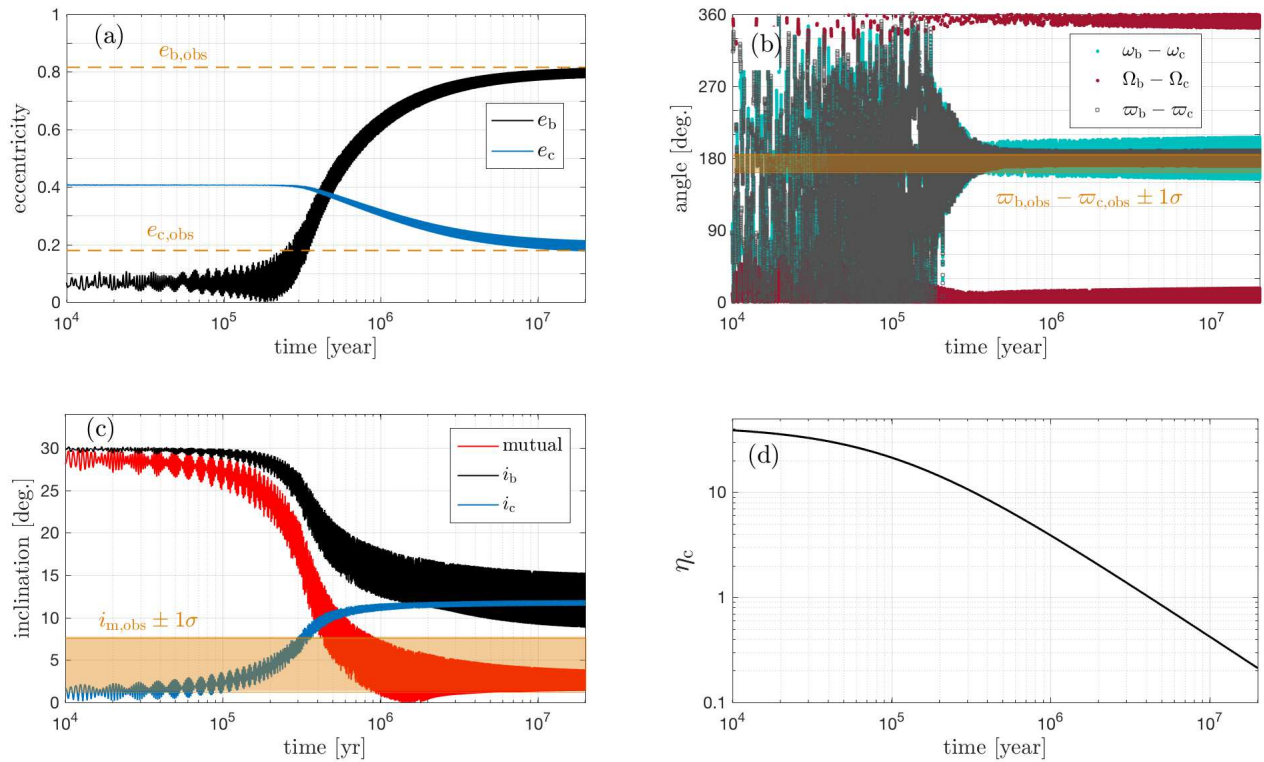}
\caption{An example system with initial parameters set to be: $e\b=0.05$, $e\c=0.4$, $i\b=30^\circ$, $i\c=1^\circ$, $\omega\b-\omega\c=150^\circ$ , and $\Omega\b-\Omega\c=60^\circ$ (implying $\varpi\b-\varpi\c=210^\circ$). The large initial value of $i\b$ is chosen to  demonstrate the effect of damping of the mutual inclinations (nodal alignment solution; upper panel of Fig. \ref{fig:i_fixed}). This example can reproduce the eccentricities, mutual inclinations and $\varpi\b-\varpi\c$  of Kepler-419 (values from Table 1 are displayed in each panel).}
\label{fig:evol_4panels}
\end{figure*}
 
 \begin{figure*}
\center
\includegraphics[width=18.5cm]{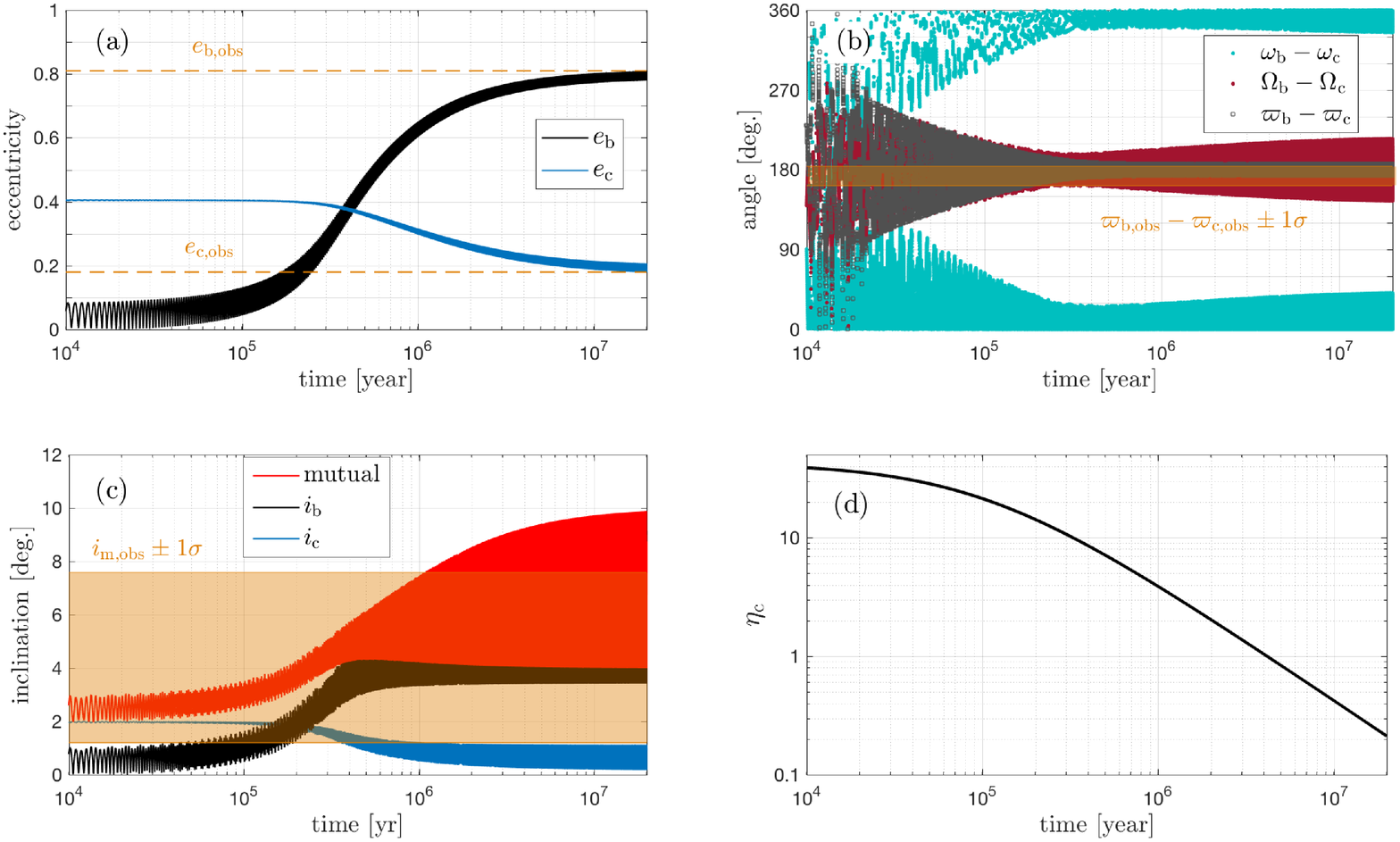}
\caption{Similar to Fig. \ref{fig:evol_4panels} but in a regime where the mutual inclinations increase. The initial parameters are: $e\b=0.05$, $e\c=0.4$, $i\b=0.1^\circ$, $i\c=2^\circ$, $\omega\b-\omega\c=60^\circ$ , and $\Omega\b-\Omega\c=60^\circ$ (implying $\varpi\b-\varpi\c=120^\circ$). The large initial value of $i\c/i\b$ leads to the increase of the mutual inclinations (nodal anti-alignment solution; upper panel of Fig. \ref{fig:i_fixed}).
 This example can reproduce the eccentricities, mutual inclinations and $\varpi\b-\varpi\c$  of Kepler-419 (values from Table 1 are displayed in each panel)}
\label{fig:anti}
\end{figure*} 

\section{Circular, mutually inclined: evolving towards co-planarity}
\label{sec:nodal}

As planets in the real system may be mutually inclined, we consider here the evolution of inclinations. To keep it simple, we address only circular orbits ($e_b = e_c = 0$).

We first need to establish a reference plane to measure the inclinations. One choice is the current invariable plane, the other is the sky-plane (observer's frame). However, the algebra is the easiest if we use the (putative) disk plane with normal $\hat{\bf n}_{\rm d}$ as reference. 

Similar to the Hamiltonian in Equation (\ref{eq:H_cop}) for eccentric orbits, we combine potentials in Equations (\ref{eq:phi_oct}) and (\ref{eq:phi_d_1.5}), to get the following dimensionless potential for inclined and circular orbits:
\ba
\tilde{\phi}_{\rm circ}&\simeq& -\eta\c {\bf \hat{j}}\c\cdot \hat{\bf n}_{\rm d}- \tfrac{1}{2}
\big({\bf \hat{j}}\b\cdot {\bf \hat{j}}\c\big)^2
\ea
or in angles (relative to $\hat{\bf n}_{\rm d}$) as
\ba
\tilde{H}_{\rm circ}&\simeq& -\eta\c \cos i_{\rm c}-\nonumber\\
&\tfrac{1}{2}&\big[\cos i\b\cos i\c+\sin i\b\sin i\c\cos\left(\Omega\b-\Omega\c\right)\big]^2.
\label{eq:H_inc}
\ea

The equations  of motion associated to $\tilde{\phi}_{\rm circ}$  can be written in a vectorial form as (see Eq. [\ref{eq:motion_tp_j}] or, e.g., \citealt{TTN09}):
\ba
\frac{d{\bf \hat{j}}\b}{d\tau}&=&\beta^{-1} ({\bf \hat{j}}\b\cdot {\bf \hat{j}}\c){\bf \hat{j}}\b\times {\bf \hat{j}}\c \\
\frac{d{\bf \hat{j}}\c}{d\tau}&=&\eta\c ~{\bf \hat{j}}\c\times \hat{\bf n}_{\rm d}+
({\bf \hat{j}}\c\cdot {\bf \hat{j}}\b){\bf \hat{j}}\c\times {\bf \hat{j}}\b,
\ea
where the dimensionless time is $\tau=t/\tau_{\rm sec,c}$ with $\tau_{\rm sec,c}$ given by Equation (\ref{eq:tau_sec}).

The equilibria of this Hamiltonian can be found in a similar way as the equilibria  from the Cassini states in the general case where the planet's spin and satellite's orbit have comparable angular momenta (e.g., \citealt{BL2006,correia2015,AL2018}).  In equilibrium, the three vectors ${\bf \hat{j}}\b$, ${\bf \hat{j}}\c$, and $\hat{\bf n}_{\rm d}$ remain in the same plane, so 
\ba
\frac{d}{dt}\left[\hat{\bf n}_{\rm d}\cdot \left({\bf \hat{j}}\b \times {\bf \hat{j}}\c\right)\right]=0,
\ea
which implies that either $\Omega\b-\Omega\c=0$ or $|\Omega\b-\Omega\c|=180^\circ$ in an inertial reference with $\hat{\bf z}=\hat{\bf n}_{\rm d}$. By replacing the equations of motion in the latter coplanarity condition, we find that the following equilibrium condition for {\it nodally aligned} planets ($\Omega\b-\Omega\c=0$) becomes
\ba
\sin \left[2(i\b-i\c)\right]\left[\frac{\sin i\c}{\beta}+\sin i\b \right]=2\eta\c \sin i\c\sin i\b 
\label{eq:Delta_Om_0}
\ea
and for {\it nodally anti-aligned} planets ($|\Omega\b-\Omega\c|=180^\circ$)
\ba
\sin [2(i\b+i\c)]\left[\frac{\sin i\c}{\beta}-\sin i\b \right]=2\eta\c \sin i\c\sin i\b.
\label{eq:Delta_Om_pi}
\ea

Furthermore, we note that equations of motion imply that the $z-$direction of the angular momentum is conserved\footnote{We have not included the back-reaction on the disk, so the total angular momentum is not necessarily conserved.}, which we express in a dimensionless form  as
\ba
\mathcal{J}_z=\beta\cos i\b+\cos i\c={\rm cst.}
\label{eq:Jz}
\ea
Thus, assuming that the planetary orbits are prograde (relative to the disk), the maximum inclination attainable by planet b is
\ba
i_{\rm b, max}=\cos^{-1} \left(\frac{\mathcal{J}_z-1}{\beta}\right).
\label{eq:ib_max}
\ea
We use this inclination as reference for Kepler-419 in this section since the values of $i\b$ and $i\c$ are unconstrained as we do not know the initial plane of the disk.

In Figure \ref{fig:i_fixed} we show the evolution of the inclinations ($i\b$, $i\c$, and mutual $i_{\rm m}$) following the fixed points as a function $\eta\c$. The upper panel  shows the fixed point at  $\Omega\b-\Omega\c=0$  (Eq. [\ref{eq:Delta_Om_0}]), which at large $\eta\c$ it corresponds to $i\b\sim i_{\rm b,max}\gg i\c$. As $\eta\c$ decreases, this fixed point leads to a decrease of the mutual inclination. On the contrary, the lower panel shows the fixed point at $|\Omega\b-\Omega\c|=180^\circ$  (Eq. [\ref{eq:Delta_Om_pi}]), which corresponds to the opposite regime where $i\b\ll i\c$ at large $\eta\c$. Here, the disk dispersal leads to an increase of the mutual inclination $i_{\rm m}$. 

The Kepler-419 system is nearly nodally aligned ($\Omega\c-\Omega\b=-5.5^\circ\pm 7.6^\circ$; Table 1) on the sky-plane reference frame. We note that this observation is a geometrical artifact of having two nearly coplanar planets where one of them transits (planet b has $i_b\simeq87^\circ$): both orbit normals lie on the same side when projected on the sky plane\footnote{\bf The mutual inclination is
given by $\cos i_{\rm m}=\cos i\b \cos i\c
+\sin i\b \sin i\c \cos(\Delta\Omega)$. If $\cos i_{\rm m}\simeq1$ and 
$\cos i\b\simeq0$, then $\sin i\c \cos(\Delta\Omega)\simeq1$, forcing 
$\Delta\Omega$ to be close to 0 (aligned nodes).}. Thus, this observation does not provide with extra dynamical information beyond the fact that the system is nearly coplanar.

In conclusion, disk dispersal can lead to either mutual inclination growth or decay in the planet pair.  If the inner planet started much more inclined than the outer one ($i\b\gg i\c$, i.e., close to the nodally aligned fixed point), the adiabatic transport will reduce the mutual inclination as the disk drains away; while if the most of the inclination is on the outer planet ($i\b\gg i\c$, i.e., close to the nodally anti-aligned mode), the mutual inclination will instead be inflated by the decaying disk. The current planets of Kepler-419 are nearly coplanar, but depending on the initial mutual inclination, they could have started from one or the other initial state.

\section{Full evolution: coupling eccentricities and inclinations}
\label{sec:evol}

Having studied the evolution in two simplified cases, here, we perform the full analysis of eccentric and inclined orbits forced by a slowly varying disk potential, using numerical integration.
It becomes clear that most of the dynamical features from our previous analysis still hold in this general case.

The equations of motion for $\evec_{\rm b}$ and $\jvec_{\rm b}$ can be written as (e.g., \citealt{M39,TTN09}):
\ba
\label{eq:motion_tp_j}
\frac{d\jvec_{\rm b}}{dt}&=&-\frac{1}{\sqrt{GM_s a\b}}\Big(
\jvec_{\rm b}\times\nabla_{\jvec_{\rm b}} \Phi
+\evec_{\rm b}\times\nabla_{\evec_{\rm b}} \Phi
\Big)\\
\label{eq:motion_tp_e}
\frac{d\evec_{\rm b}}{dt}&=&-\frac{1}{\sqrt{GM_s a\b}}\Big(
\jvec_{\rm b}\times\nabla_{\evec_{\rm b}} \Phi
+\evec_{\rm b}\times\nabla_{\jvec_{\rm b}} \Phi
\Big),
\ea
where $\Phi=\phi_{\rm planet}+\phi_{\rm disk}$ (Eqs. [\ref{eq:phi_oct}] and [\ref{eq:phi_d_1.5}]), with corresponding
equations for $\evec_{\rm c}$ and $\jvec_{\rm c}$ (swap sub-indices b for c).

We assume the gas disk has an inner cavity with the inner edge  at $R\in=1.5a\c\sim2.5$ AU, and has a radial extent of $R_{\rm out}/R_{\rm in}=10$, with a  power-law index $\gamma=3/2$ for the density profile (Eq. [\ref{eq:sigma_r}]). We adopt an initial disk mass of $0.1 M_s$ and let this decay with time as $\propto1/(1+t/\tau_{\rm v})$ \citep[see, e.g.,][]{BA2013}, 
\ba
\eta_c(t) \simeq\frac{2.1M_{\rm disk}}{m_{\rm c}}=\frac{40}{1+t/\tau_{\rm v}}\, .
\ea
Here, we adopt a decay time of  $\tau_{\rm v}=10^{5}$ yr. 
While longer values of $\tau_{\rm v}$ are possibly more realistic, our choice already allows the adiabatic approximation to be valid and should not alter the dynamics (while it speeds up the numerical integration).

We first show two cases that have specifically chosen initial parameters and that successfully evolve into Kepler-419 today, before discussing all possible initial states for Kepler-419.

Both cases start with most of the AMD initially deposited in the outer planet, and the inner planet is nearly circular.   As is discussed in \S \ref{sec:apsidal}, as long as the initial $\eta_c$ is sufficiently large, we expect eccentricity to be gradually transferred to the inner one as the disk drains away, regardless of the initial apse orientations. The inclinations require a bit more care. The first case (Fig. \ref{fig:evol_4panels}) is initialized such that, the inner orbit is much more inclined than the outer one ($i\b=30^\circ$ and $i\c=1^\circ$).
The reverse is true for the case in Figure \ref{fig:anti} where $i\b=0.1^\circ$ and $i\c=2^\circ$. According to the analysis in \S \ref{sec:nodal}, mutual inclination decays in the first case, and grows in the second case.
We have chosen the initial mutual inclination accordingly, in order to reproduce today's value (nearly coplanar). 

In conclusion, results from 3-D numerical integration exhibit much of the same features as we have discussed in the simple case, in particular, features like eccentricity transfer to the inner planet and final apse anti-alignment seem unaffected by the inclusion of moderate inclinations.

\subsection{Generality of Initial conditions to match Kepler-419}

Here, we study the likelihood that a two-planet system would have evolved to the configuration we see today in Kepler-419.
Instead of running a full population synthesis, we rely on the principle of adiabatic invariance and integrate backward in time, to put constraints on what the initial values of $e\b$ (and, therefore, $e\c$), and $\varpi\b -\varpi\c$ can reproduce the system. 

\begin{figure*}
\center
\includegraphics[width=16cm]{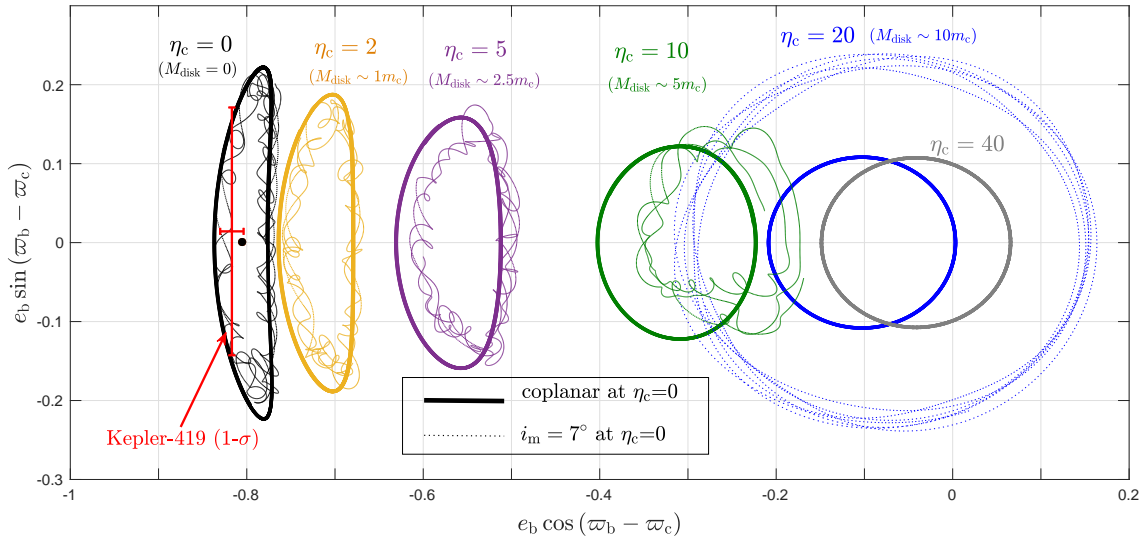}
\caption{Possible initial conditions for the Kepler-419 system. The left-most thick black ellipse encompasses all possible trajectories that fall within $1-\sigma$ measured values of the Kepler-419 pair, assuming they are coplanar. As the value of $\eta_c$ (disk precession) rises successively, this ellipse transforms to thick ellipses to the right, together with all its internal trajectories. This therefore yields the possible initial condition at early times. The thin dotted lines illustrate the more complex situation when the planets are not coplanar (see text for explanations).
}
\label{fig:evol_lib90}
\end{figure*}

 We first investigate the case of a coplanar pair that also lie in the disk plane. To obtain possible initial conditions, we place the pair on a current trajectory ($\eta\c = 0$) that encompasses the $1-\sigma$ error bar of Kepler-419. This is shown as the black ellipse (thick line) to the left of Figure \ref{fig:evol_lib90}. This is the trajectory with the largest libration amplitude (within error). We then integrate this trajectory backward in time, increasing $\eta\c$ gradually. Its equivalent trajectories at different values of $\eta_c$ are shown as ellipses in other colors. By the time $\eta\c \sim 40$, the last ellipse becomes a trajectory that circulates around  the fixed point close to the origin. As adiabatic transport conserve the volume of phase space, any trajectory initially falling within the last ellipse, going forward in time, will automatically fall within the first ellipse, and is therefore compatible with the observed state of Kepler-419. In other words, for the coplanar case, as long as $\eta_c$ was sufficiently high in the past ($\geq 20-40$), the appropriate initial conditions are those with $e_b \lesssim 0.1$ and arbitrary apse angle ($|\varpi\b - \varpi\c|$ unconstrained). This is generic.

The possible set of initial conditions for the general non-coplanar case is more complex. We assume a current mutual inclination of $7^\circ$ (close to $1-\sigma$ upper limit), by setting $i\b=12^\circ$, $i\c=5^\circ$, and $\Omega\b - \Omega\c = 0$. This is measured in the original disk plane. We plot the results of the backward integration as thin dotted lines in Fig. \ref{fig:evol_lib90}, up to $\eta_c = 20$. 
They look largely similar to the coplanar case at most values of $\eta_c$, with a roughly constant phase space volume in the plotted plane, and encompass circulating trajectories by the time $\eta_c = 20$.  We note that although this set of initial conditions is arbitrary, our choice enhances the effect from departure from coplanarity because the mutual inclinations increase as we integrate backwards in time (nodally aligned evolution mode, similar evolution to that in figure \ref{fig:evol_4panels}).

At large $\eta_c$, area in this projection of phase-space appears to increase, at the expense of a decrease in the volume of the other dimensions (phase-space volume $\propto d\Omega\b d\Omega\c  i\b^2 i\c^2$).
Similarly, as trajectories travel in a higher dimensional phase space, it is no longer true that all points falling within the initial trajectory (in the plotted plane) will remain so for all times. So not all initial conditions that fall within the thin dotted curve at large $\eta_c$ is guaranteed to lie close to the observed system today. But one can roughly conclude that 
for $\eta\c\gtrsim20$, the range of apsidal orientations that match the observation is large if $e\b\lesssim0.2$.

In conclusion, as long as the initial $e_b \lesssim 0.1$ and the initial $\eta_c \gtrsim 20-40$, we will likely obtain a system like Kepler-419 today.\footnote{There is an implicit condition that the initial AMD is the same as today. This requires $e\c \geq 0.4$.}
 The latter condition translates to an initial disk mass  $M_{\rm disk} \gtrsim 10 m_c \simeq 0.05 M_s$.


\section{Discussions}
\label{sec:discussion}

We have  demonstrated that the secular orbital dynamics of a two-planet system, subject to the gravitational perturbations from an axisymmetric protoplanetary disk that gradually disperses, will naturally land the planets into the current observed state of Kepler-419, where the two planets have anti-aligned apses and the inner one obtains a very high eccentricity. For ease of discussion, we term this process `secular transport'.

In what follows, we discuss our model assumptions and  implications of secular transport for other exoplanet systems, as well as its relation to previous work.

\subsection{Previous work}

The secular dynamics of a coplanar two-planet system influenced by a decaying disk have been previously studied by \citet{naga2003}. Consistent with our results, the authors note that AMD can be efficiently transferred between orbits and could lead to an imprint on the relative apsidal orientations.\footnote{\citet{naga2003} called this 'sweeping' of `secular resonance'. We differ subtly in our interpretation.} 

The authors focused on the $\upsilon$ Andromedae system trying to explain its apsidal alignment ($\omega\c-\omega_{\rm d}\sim0$ measured from RV, not $\varpi\c-\varpi_{\rm d}$ ). There are a few short-comings with this system: (i) astrometric constraints place the planets in mutually inclined orbits ($\simeq30^\circ$) \citep{McArthur2010}; (ii) the astrometric constraints on $\Omega\c-\Omega_{\rm d}$ place the systems far from the aligned fixed point, closer to the anti-aligned one \citep{Deitrick2015};
(iii) the planet-planet scattering model has no major difficulty on reproducing the secular state of the system \citep{barnes2011}.  
Thus, $\upsilon$ Andromedae, as well as other systems explored by \citet{naga2003}, do not provide unambiguous evidence of being sculpted during the disk dispersal phase.

Unlike $\upsilon$ Andromedae, the orbital parameters of Kepler-419 robustly place the orbits close to a secular fixed point \citep{alme2018}, which is hard to be populated by planet-planet scattering \citep{BG06,BG07}. The coplanarity also argues against scattering \citep{chatt2008}.

\subsection{Model Assumptions}

Our model requires the two following key ingredients: first, there is a disk that causes rapid precession on the outer planet\footnote{We modeled a disk with a cavity causing faster (prograde) precession of the outer planet. The dynamics is essentially unchanged if the inner planet precesses faster than the outer one, but in a retrograde sense. The model requires a disk that drives an effective retrograde precession, $\dot{\varpi}\b-\dot{\varpi}\c<0$.}; second, there is an initial amount of AMD. The latter is conserved in our model (which includes only secular interactions) and is determined by the current observed value.

Our model only considers the precessional effect of the disk on the planets. To be able to ignore other effects (e.g., eccentricity damping, inclination damping), it is preferable if this disk lies outside planet c. We therefore posit a disk that has a large inner cavity, with an inner edge that lies at $R_{\rm in} \sim 1.5 a_c \sim 2.5$ AU. This may be compatible with the population of the so-called `transitional disks' \citep{espaillat}. The clearing of the inner region, in this case, may be performed by the two giant planets\footnote{In particular, we note that the outer planet is assumed to have an apoapse that lies around $a_c(1+e_c)  \sim 2.4$ AU. This will cause additional clearing over models assuming circular planets.} \citep[e.g.,][]{goldreich,crida2006,kanagawa,zhu2011}. Photoevaporation also provides another plausible way of clearing out the inner region \citep{uvswitch,owen2010,owen2011,reviewpe}. Stellar winds \citep{rt1,rt2} may also remove angular momentum of the inner disk and carve a large hole.

For such an external disk, our model shows that a relatively large disk mass is required, if the initial apse angles are randomly distributed (Fig. \ref{fig:evol_lib90}). The condition that $\eta_c \geq 20-40$ is translated into $M_{\rm disk} \gtrsim 0.05 - 0.1 M_s$. While these values are on the higher end of the observed protoplanetary disks mass distribution, such massive disks do seem to exist according to infrared and sub-mm surveys with \textit{Herschel} and ALMA \citep{mcclure, pasc2,long2}.

We now turn to consider origin for the initial AMD required to explain Kepler-419. This should be in the form of a moderate $e_c$. According to current understanding, this can arise from planet-disk interactions, and/or from planet-planet scatterings.
Regardless of the AMD origin, we note that giant planets detected in the radial velocity surveys commonly exhibit large eccentricities, suggesting that much AMD is universally present in these systems.
 
\paragraph{Disk-planet interactions}
Eccentricity in planet c can be excited by  gravitational torques from the outer disk  \citep[e.g.,][]{goldreich,GS2003,OL2003}.
For planets massive enough to open a wide gap, as likely is the case for planet c,  the co-rotation torques, which generally damp the eccentricities \citep{GS2003,OL2003}, can be sharply reduced and  contribution from the outer Lindblad resonance ($1:3$) may dominate. The latter  is known to  excite planet eccentricity
(e.g., \citealt{lubow1991,pap2001,bitsch2013,dunhill2013}).  We note that there are still substantial theoretical uncertainties in this scenario and it is a topic of ongoing research (e.g., \citealt{duffel2015,rosotti2017,Ragusa2018}).

Related to the AMD injection by the disk,  \citet{CM2002} showed that this process could by itself lead to the apsidal locking of two secularly-interacting planets (e.g., $\upsilon$ Andromedae). Consistent with their results, we experimented by driving $e\c$ to its current value of  and ignoring the disk-driven precession,  and found apsidal locking around alignment. This is contrary to the current state of Kepler-419. However, we found that we driving  $e\c$  from 0 to $0.17$ and include disk-driven precession (i.e., the adiabatic transport studied here), we find that the 3-D orbital architecture Kepler-419 can be fully explained. This process does not require the disk to excite $e_c$ up to 0.4, but up to 0.17,  and deserves further study.

\paragraph{Planet-planet scattering}
Another possibility is that the AMD results from planet-planet gravitational scattering in the disk phase \citep[e.g.,][]{marzari2010,moeckel2012,lega2013}.  For instance, hydrodynamic simulations by \citet{lega2013}   that include N-body interactions between multiple giant planets show that  planetary orbits can frequently become  destabilized by the action of the gas disk, with subsequent planet-planet scatterings leading to excitation of both eccentricities and inclinations (e.g., \citealt{chatt2008,JT2008,lega2013}).
If large inclinations are excited, we showed that the adiabatic transport can damp the mutual inclinations to match current state of Kepler-419 (Figure \ref{fig:evol_4panels} starting from $\simeq30^\circ$).

\subsection{Implications for other exoplanet systems}

One of the key features of the secular transport process  is the production of large eccentricities and (possibly) inclinations in systems with only two planets. This can have important implications for different populations of exoplanets as we describe next.

\paragraph{Eccentric warm Jupiters with outer companions.}
    Radial velocity surveys have disclosed a population of warm Jupiters (with periods between $10$ to $100$ days)  that have outer cold Jupiter companions.  Kepler-419 may well belong to this class. Interestingly, these warm Jupiters are significantly more eccentric than other warm Jupiters lacking  such companions \citep{DC2014,dong2014,PT2016,bryan2016}.  Their eccentricities are not easily explained by planet-planet interactions as scattering becomes inefficient at exciting eccentricities at these close-in distances \citep{petro14} and the outer companions are too far away for scattering,  leaving secular interactions as a more likely candidate. While secular interactions within a highly mutually inclined pair can explain these eccentricities by the Lidov-Kozai mechanism \citep{AL2017}, another intriguing possibility is that these warm Jupiters acquired their eccentricities by  adiabatic transport during the disk-clearing stage, much like the scenario we discuss here for Kepler-419.

A related puzzle, as is  pointed out by \citet{DC2014}, is that these eccentric warm Jupiters with companions, unlike Kepler-419, tend to have perpendicular apses\footnote{For these planets, unlike Kepler-419, there are no constraints on the viewing angle or mutual inclinations, and $\Delta\omega$ on the plane of the sky determined from radial velocity measurements is used as a proxy for $\Delta\varpi$ measured in the invariable plane.} ($|\varpi\b-\varpi\c|\sim90^\circ$). At face value this trend would argue against our model which favors anti-alignment. However, as  \citet{DC2014} argued, these planets may be librating around $|\varpi\b-\varpi\c|\sim180^\circ$ with large amplitudes.  Such a pattern may occur when the mutual inclinations are high and we plan to explore the detailed dynamics in a subsequent work.

\paragraph{Eccentric cold Jupiters}
Many of the cold Jupiters (outside $100$ days) orbit with high eccentricities. Although planet-planet scattering and planet-disk interactions may possibly give rise to eccentric orbits, it is possible that some of the very eccentric cold Jupiters are produced in the disk clearing stage. In this paper, we show that extreme eccentricities may be produced in the presence of an outer companion (that is moderately eccentric) and a decaying disk.

The following application considers the extreme case when planet eccentricities are excited to such high values that they start tidal interaction with the central stars.

\paragraph{Hot Jupiters}
 Roughly half of the hot Jupiters are suggested to have distant planetary-mass companions \citep{knutson2014,bryan2016}.  If these planets reached their current orbits by high-eccentricity migration (see \citealt{DJ2018} for a recent review), adiabatic transport during disk dispersal may provide a novel venue to excite their eccentricities (and possibly inclinations) to very high values.

We note that the excitation of extreme eccentricities and inclinations in the context of a two-planet system perturbed by an external disk has been studied by \citet{chen2013}. The authors focused on the quadrupole-level Lidov-Kozai mechanism for a highly inclined two-planet system and perturbations from a non-dispersing gas disk.  Our work emphasizes the relevance of two extra ingredients missing in their study: (i) octupolar planet-planet interactions; (ii) gradual removal of the gas disk. These ingredients would allow for eccentricity growth and migration even if the planets have small mutual inclinations.  In Kepler-419, the inner planet reaches an eccentricity of $0.82$--this value can be further increased for favourable system parameters. 
For example, if the mass of Kepler-419b is decreased from the current value of $\simeq 2.8M_J$ to $1M_J$, then the fixed point in the coplanar model moves from $e\simeq0.82$ to $e\simeq0.92$ (Eq. [\ref{eq:fixed_point}]), so its pericenter distance decreases from $\simeq0.07$ AU to $\simeq0.03$ AU, possibly allowing for migration to a hot Jupiter orbit.

Effects beyond the simple coplanar model, like excitation of large inclinations, large libration amplitudes around the fixed point, possible separatrix-crossing events when short-range forces are included \citep{MG2009}, and other effects might play a significant role at determining the maximum eccentricities that can be reached during disk dispersal as well the range of inclinations. We plan to explore these effects in a subsequent work.

\section{Conclusions}

We have studied the secular gravitational coupling between two planets as their birth protoplanetary disk gradually disperses, using the Kepler-419 system as an example. We show that when the precession of the planets is largely dominated by the disk, the planets circle around an anti-aligned secular state. As the disk depletes, the system can be adiabatically transported to a state where most of the eccentricity (or rather, the angular momentum deficit) is on the inner orbit and the relative apsidal orientations are largely anti-aligned. The inclinations are subject to similar evolution.

We refer to this process as  `adiabatic transport' and we show that it naturally explains the intriguing orbital architecture of the Kepler-419 system, which has two nearly coplanar giant planets with large eccentricities ($e\b\simeq0.82$) and relative apsidal orientations librating around anti-alignment. Other proposals to achieve these large eccentricities are unlikely to explain its orbital state. Our model requires that the disk had a mass of at least several percent that of the host star at the moment that the regions inside $\sim2$ AU were largely depleted. 
 
We argue that the mechanism studied here may be important for forming hot Jupiters
and explaining the large eccentricities of some warm and cold Jupiters.\\

\acknowledgements 
We thank the referee for useful and thoughtful comments. We also thank  Adrian Hamers, Chris Spalding, Dan Tamayo, Diego Mu\~noz, Dong Lai,  Hilke Schlichting, Konstantin Batygin, Man Hoi Lee, Roman Rafikov, and Scott Tremaine for comments on an early version of this manuscript. CP acknowledges support from the Gruber Foundation Fellowship and Jeffrey L. Bishop Fellowship. YW thanks NSERC for research support.

\appendix
\section{Orbit-averaged potential from an external disk}

We consider an axisymmetric disk with a truncated density profile
\ba
\Sigma(R)=\Sigma_0 \left(\frac{R}{R\in}\right)^{-\gamma}~~ \mbox{for}~~R\in<R<R\out,
\ea
and $\Sigma(R)=0$ elsewhere.
The disk has a mass
\ba
M_{\rm disk}=\int_{R\in}^{R\out}\Sigma_0\left(\frac{R\in}{R}\right)^{\gamma}2\pi RdR=2\pi\Sigma_0 R\in^2\left(\frac{\left(R\out/R\in\right)^{2-\gamma}-1}{2-\gamma}\right).
\ea

\begin{figure}
\center
\includegraphics[width=9.3cm]{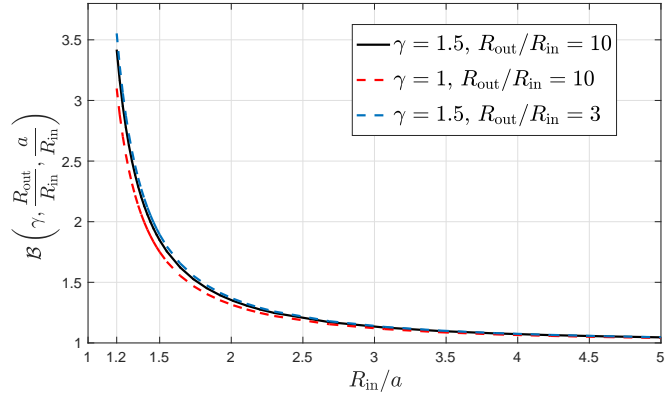}
\caption{Factor in Equation (\ref{eq:B}) to account for the contribution of the disk potential from regions close to the planet that are not properly described for the limiting potential in Equation (\ref{eq:phid_disk_limit}) for distant disks ($a_p\ll R\in$). As expected the corrections are significant only when $a\lesssim R\in$ and it depends weakly on the density profile slope $\gamma$
and the radial extent $R\out/R\in$.}
\label{fig:beta}
\end{figure}

We construct the potential from the disk taking annuli of mass
$dm_{\rm disk}(R)=2\pi \Sigma(R) RdR$ and integrate this over the whole disk. 
This procedure is similar to the previous work by \citet{ward81}, which has been extended to eccentric disks by \citet{SR2015} and \citet{ST2018}, and shown  to reproduce the precession rates of opening-gap planets from hydrodynamical simulations by \citet{FM2016}. 
The potential for a ring with zero eccentricity and normal $\hat{\bf n}_{\rm d}$ averaged over the orbit of a planet with mass $m_p$ and semi-major axis $a_p$ can be written to second order in eccentricities and mutual inclinations as (e.g., \citealt{BF2014})
\ba
d\phi_{\rm disk}=-\frac{Gm_p dm_{\rm disk}(R) a_p}{R^2}\cdot\tfrac{1}{4} b_{3/2}^{(1)}(a_p/R)
\left[e_p^2+ {\bf j}_p\cdot \hat{\bf n}_{\rm d}\right]+\mbox{ cst.},
\ea
where the Laplace coefficient is defined as
\ba
b_{3/2}^{(1)}(\alpha)=\frac{1}{\pi}\int_0^{2\pi}\frac{\cos (\phi)d\phi}{(1-2\alpha\cos\phi+\alpha^2)^{3/2}}.
\ea
For $\alpha\ll1$, $b_{3/2}^{(1)}(\alpha)\simeq 3\alpha$, so the potential for
$a_p\ll R\in$ can be written as
\ba
\phi_{\rm disk}\big |_{a_p\ll R\in}=\int_{R\in}^{R\out}d\phi_{\rm disk}\big|_{a_p\ll R\in}\to-\frac{3Gm_p M_{\rm disk} a_p^2}{4R\in^3}K\left(\gamma,\frac{R\out}{R\in}\right)
\left[e_p^2+ {\bf j}_p\cdot \hat{\bf n}_{\rm d}\right]+\mbox{ cst.},
\label{eq:phid_disk_limit}
\ea
where
\ba
K\left(\gamma,\frac{R\out}{R\in}\right)=\frac{2-\gamma}{\gamma+1}\left(\frac{1-\left(R\out/R\in\right)^{-1-\gamma}}
{\left(R\out/R\in\right)^{2-\gamma}-1}\right).
\ea
A similar expression was found by \citet{Terquem2010}.
Since the inner edge of the gap can be close to the outer planet, we write the potential from integrating Equation (\ref{eq:phid_disk_limit}) as
\ba
\phi_{\rm disk}=-\frac{3Gm_p M_{\rm disk} a_p^2}{4R\in^3}K\left(\gamma,\frac{R\out}{R\in}\right) 
\mathcal{B}\left(\gamma,\frac{R\out}{R\in},\frac{a_p}{R\in}\right)
\left[e_p^2+ {\bf j}_p\cdot \hat{\bf n}_{\rm d}\right]+\mbox{ cst.},
\label{eq:phi_d}
\ea
where we introduce
\ba
\mathcal{B}\left(\gamma,\frac{R\out}{R\in},\frac{a_p}{R\in}\right)
\equiv \frac{1}{3} \left(\frac{R\in}{a_p}\right)\left(\frac{\gamma+1}{1-\left(R\out/R\in\right)^{-1-\gamma}}\right)
\int_1^{R\out/R\in} u^{-\gamma} b_{3/2}^{(1)}\left(\frac{a_p}{R\in}\frac{1}{u}\right)du,
\label{eq:B}
\ea
as a correction factor for contribution of the rings close to $a$ for which the limiting polynomial expression when $b_{3/2}^{(1)}(\alpha)\to 3\alpha$.
The function $\mathcal{B}$ is shown Figure \ref{fig:beta}, where we observe that it mainly depends on $a_p/R\in$, rapidly increasing as it approaches unity, as expected. We note, however, that there is no real divergence of the potential as the disk starts embedding the planets and we only use our results only for $R\in/a_p\gtrsim1.5$.

\end{document}